\documentclass[11pt,english]{amsart}

\usepackage{amsfonts,amsmath,latexsym,verbatim,amscd,color,array}

\usepackage[colorlinks=true]{hyperref}

\newtheorem{theorem}{Theorem}[section]
\newtheorem{proposition}{Proposition}[section]
\newtheorem{lemma}{Lemma}[section]
\newtheorem{definition}{Definition}[section]
\newtheorem{corollary}{Corollary}[section]
\newtheorem{remark}{Remark}[section]
\newtheorem{example}{Example}[section]

\title{A note on some overdetermined elliptic problem}

\author{Fr\'ed\'eric H\'elein}
\address{Institut de Math\'ematiques de Jussieu, UMR CNRS 7586, 
Universit\'e Paris Diderot--Paris 7, Case
7012, B{\^a}timent Chevaleret, 75205 Paris Cedex 13, France}
\email{helein@math.jussieu.fr}

\author{Laurent Hauswirth}
\address{Universit\'e de Marne la Valle, Cit\'e Descartes - 5 boulevard Descartes
Champs-sur-Marne, 77454 Marne-la-Vall\'ee Cedex 2, France}
\email{Laurent.Hauswirth@univ-mlv.fr}

\author{Frank Pacard}
\address{Universit\'e Paris Est-Cr\'eteil et Institut Universitaire de France, 61 Avenue du 
G\'en\'eral de Gaulle, 94010 Cr\'eteil, France}
\email{pacard@univ-paris12.fr}
\thanks{Acknowledgments : We thank Romain Dujardin and Charles Favre for useful discussions. The third author is also partially supported by the ANR-08-BLANC-0335-01 grant.}

\begin{document}

\maketitle

\section{Introduction}
\setcounter{equation}{0}

Given $(M,g)$, a $m$-dimensional Riemannian manifold, and $\Omega$, a smooth 
bounded domain in $M$, we denote by $\lambda_1 (\Omega)$ the first eigenvalue of 
the Laplace-Beltrami operator under $0$ Dirichlet boundary condition. The critical 
points of the functional
\[
\Omega \longmapsto \lambda_1 (\Omega) \, ,
\]
under the volume constraint $\mbox{Vol} (\Omega) = \alpha$ (where  $\alpha \in (0,
\mbox{Vol} (M))$ is fixed) are called {\em extremal domains}. Smooth {\em 
extremal domains} are characterized by the property that the eigenfunctions 
associated to the first eigenvalue of the Laplace-Beltrami operator have constant 
Neumann boundary data \cite{ElS-Il}. In other 
words, a smooth domain is extremal if and only if there exists a positive function 
$u_1$ and a constant $\lambda_1$ such that 
\[
\Delta_g u_1 + \lambda_1 \, u_1 = 0 \, ,
\]
in $\Omega$ with 
\[
u_1 =0 \qquad \mbox{and} \qquad \nabla_n u_1 = {\rm constant} \qquad \mbox{on} 
\qquad \partial \Omega \, ,
\]
where $n$ denotes the inward unit normal vector to $\partial \Omega$.
The theory of {\em extremal domains} is very reminiscent of the theory of constant mean 
curvature surfaces or hypersurfaces. To give some credit to this assertion, let us
recall that, in the early 1970's, J. Serrin has proved that the only 
compact, smooth, extremal domains in Euclidean space are round balls  \cite{Ser}, 
paralleling the well known result of Alexandrov asserting that round spheres are the 
only (embedded) compact constant mean curvature hypersurfaces in Euclidean space. 
More recently, F. Pacard and P. Sicbaldi have proved the existence of extremal domains 
close to small geodesic balls centered at  critical points of the scalar curvature function 
\cite{Pac-Sci}, paralleling an earlier result of R. Ye which provides constant mean 
curvature topological spheres (with high mean curvature) close to small geodesic 
spheres centered at nondegenerate critical points of the scalar curvature function \cite{Ye}.
 
We propose the following~:
\begin{definition}
A  smooth domain $\Omega \subset \mathbb R^m$ is said to be an {\em exceptional 
domain} if it supports positive harmonic functions having $0$ Dirichlet boundary data 
and constant (nonzero) Neumann boundary data. Any such harmonic function is called 
a \emph{roof} function.
\end{definition}

{\em Exceptional domains} arise as limits under scaling of sequences of {\em extremal 
domains} just like minimal surfaces arise as limits under scaling of sequences of constant 
mean curvature surfaces. As explained above, there is a formal correspondence between 
{\em extremal domains} and constant mean curvature surfaces. In this note, we try
to explain that  there is also a strong analogy between {\em exceptional domains} and 
minimal surfaces.  More generally, we propose the~:
\begin{definition}
A  $m$-dimensional flat Riemannian manifold $M$ is said to be {\em exceptional} 
if it supports positive harmonic functions having $0$ Dirichlet boundary data and constant 
(nonzero) Neumann boundary data. Any such harmonic function is called a 
{\em roof} function.
\end{definition}

Our results raise the problem of the classification of (unbounded) smooth $m$-dimensional 
{\em exceptional manifolds}. In trying to address this classification problem, we provide a 
Weierstrass type representation  formula for {\em exceptional flat surfaces}. 
When the dimension $m=2$, we give non trivial examples of {\em exceptional domains} 
which are embedded in $\mathbb R^2$ and we prove a {\it half space} result for {\em
exceptional domains} which are conformal to a half plane. 

\section{A non trivial example of {\em exceptional domain} in $\mathbb R^2$}
\setcounter{equation}{0}

To begin with, observe that the property of being an {\em exceptional domain} is preserved under the 
action of the group of similarities of $\mathbb R^m$ (generated by isometries and dilations). 
We now give trivial examples of {\em exceptional domains} in $\mathbb R^m$~: 

\begin{itemize}

\item[(i)] The half space $\{ x = (x_1, \ldots, x_m) \in \mathbb R^m \, : \, x_1 >0 \}$ is an {\em exceptional 
domain} in $\mathbb R^m$ since the function $u(x) = x_1$ is a positive harmonic function with 
$0$ Dirichlet boundary data and constant Neumann boundary data. \\

\item[(ii)]  The complement of a ball of radius $1$ in $\mathbb R^m$ is an {\em exceptional 
domain} since, the function $u$ defined by $u (x) : = \log | x |$, when $m=2$ and  
$ u(x) : = 1 - |x|^{2-m}$, when $m \geq 3$ is positive, harmonic and has $0$ Dirichlet and  
constant Neumann data on the unit sphere. \\

\item[(iii)] The product $\Omega \times \mathbb R^k$ is an {\em exceptional domain} in $\mathbb R^m$ 
provided $\Omega \subset \mathbb R^{m-k}$ is an {\em exceptional domain} in $\mathbb R^{m-k}$.
\end{itemize}

In dimension $m=2$, there exists (up to a similarity) at least another {\it exceptional 
domain}. To describe this domain, we make use of the invariance of the Laplace 
operator under conformal transformations. The idea is that there exists a (somehow natural) 
unbounded, positive harmonic function $U$ with $0$ Dirichlet boundary condition on 
an infinite strip in $\mathbb R^2$. This function does not have constant Neumann 
data but we can then look for a conformal transformation $h$ which has the property 
that the pull back of the harmonic function $U$ by  $h$ has constant Neumann boundary 
data on the boundary of the image of the strip by $h$. 

To proceed, it is be convenient to identify $\mathbb R^2$ with the complex plane 
$\mathbb C$. We claim that~:
\begin{proposition}
The domain 
\[
\Omega : = \left\{ w \in \mathbb C \, : \,  |\Im \, w| < \tfrac{\pi}{2} + \cosh (\Re \, w)  \right\} \, ,
\]
is an {\em exceptional domain}.
\label{pr:2.1}
\end{proposition}

To prove this result, we define the infinite strip
\[
S : =  \left\{ z  \in \mathbb C  \, : \,    \Im \, z  \in (- \tfrac{\pi}{2} , \tfrac{\pi}{2} ) \right\} \, ,
\]
and the mapping 
\[
F (z) : =   z +  \sinh z \, . 
\]
Observe that $\Omega =  F  (S)$. The proof of Proposition~\ref{pr:2.1} follows from the following two results.
\begin{lemma}
The mapping $F$ is a conformal diffeomorphism from $S$ into $\Omega$. 
\end{lemma}
\begin{proof}
We can write
\[
F (z) - F (z') =  (z-z')  \, \int_0^1 ( 1+  \cosh \, (t z+ (1-t) z') )\, \, dt  \, .
\] 
In particular 
\begin{equation}
\langle z-z', F (z) - F (z') \rangle = | z - z ' |^2 \, \left( 1+  \int_0^1 \, \Re \, 
\cosh \, (t z+ (1-t) z') \, dt \right)  \, ,
\label{eq:inj}
\end{equation}
where $\langle \cdot , \cdot \rangle$ denotes the scalar product in $\mathbb C$. 
Now, observe that, for all $x+i\, y \in S$, we have 
\[
\Re \, \cosh  \, (x+i \, y) = \cosh x \, \cos y  \geq 0 \, .
\]
This, together with (\ref{eq:inj}), implies immediately that $F$, restricted to $S$, is
injective.  We also have 
\[
|\partial_z \Lambda  (z)|^2 = |1+ \cosh z|^2 = ( \cosh x + \cos y )^2 \, .
\] 
Therefore, $\partial_z F $ does not vanish in $S$ and this shows that $F$ 
is a local diffeomorphism and the mapping $F$ being holomorphic, it is 
conformal. \end{proof}

We define the real valued function $u$ on $\Omega$ by the identity
\[
 u ( F (z) ) = \Re  \, \cosh z  \, ,
\]
for all $z \in S$.  We have the~:
\begin{lemma}
The function $u$ is harmonic and positive in $\Omega$, vanishes and has constant 
Neumann boundary data on $\partial \Omega$.
\end{lemma}
\begin{proof}
The function $W$ defined in $\mathbb C$ by $W (z) : = \Re \, \cosh z$ 
is harmonic. As already mentioned in the proof of the previous Lemma, 
$W (x+i \, y) = \cosh x\, \cos y$  and hence, the function $W$ is both harmonic 
and positive in $S$ and vanishes on  $\partial S$. The mapping $F$ being 
a conformal diffeomorphism from $S$ to $\Omega$, we conclude the function 
$u$ is both harmonic and positive in $\Omega$ and vanishes on $\partial \Omega$. 
We claim that $u$ has constant Neumann data on $\partial \Omega$. Indeed, by
definition
\[
u ( F  (z))  = \tfrac{1}{2} \, ( \cosh z + \cosh \bar z) \, .
\]
Since $F $ is holomorphic, differentiation with respect to $z$ yields  
\[
2 \,  \partial_z u  ( F (z)) = \frac{ \sinh z }{1+ \cosh z} \, .
\]
Therefore
\[
|\nabla u|^2  (F (z)) =  \frac{\cosh x - \cos y}{\cosh x+ \cos y} \, ,
\]
where $z = x+ i \, y$.  On $\partial \Omega$,   $y = \pm \pi/2$ and 
hence $| \nabla u|  \equiv 1$. Since we already know that $u =0$ on 
$\partial \Omega$, we conclude that  $u$ has constant Neumann boundary data.
\end{proof}

The two previous Lemma complete the proof of the fact that $\Omega = F (S)$
 is an {\em exceptional domain} in $\mathbb R^2$ with {\em roof} function given by $u$. 
 
 \begin{remark}
We suspect that this example generalises to any dimension $m \geq 3$, namely that 
there exists a rotationally symmetric  {\em exceptional domain} in $\mathbb R^m$, for 
all $m \geq 3$. 
 \end{remark}

\section{Toward a global representation formula}\label{preglobal}
\setcounter{equation}{0}

Let $M$ be a {\em exceptional flat surface} with smooth boundary $\partial M$. 
Let $\tilde{M}$ be its universal cover and $\partial \tilde{M}$ be the preimage 
of $\partial M$ by the covering map $\tilde{M} \longrightarrow M$. In the 
following, we exclude the non interesting case where $\partial M = \emptyset$.

By assumption, $M$ is a flat surface and hence $\tilde{M}$ is naturally endowed 
with a flat Riemannian metric $g$ and hence with an induced complex structure 
which is conformal to the standard one. Also,  there exists an orientation 
preserving isometric immersion $F: (\tilde{M},g)\longrightarrow (\mathbb{C},
g_\mathbb{C})$ (where $g_\mathbb{C}$ is the canonical Euclidean metric 
on $\mathbb{C}$) which induces a smooth immersion of $\partial \tilde{M}$. 
Observe that $F$ is holomorphic and that 
\[
\|dF\|_g = 1\, ,
\]
in $\tilde M \cup \partial \tilde{M}$. We define the holomorphic $(1,0)$-form
\[
\Phi:= dF  = \partial_z F \, dz \, ,
\]
Observe that $\Phi$ does not vanish and admits  a smooth extension to 
$\tilde{M}\cup\partial \tilde{M}$.

We let $u: M\longrightarrow \mathbb R^+$ be a {\em roof}  function on $M$ 
and, with slight abuse of notation, we denote also by $u: \tilde{M} \longrightarrow 
\mathbb R^+$  its lift. The {\em roof} function $u$ can be normalized so that 
\begin{equation}
\| \nabla u \|_g = 1 \, ,
\label{eq:nblu1}
\end{equation}
on $\partial M$. We consider the harmonic conjugate function 
$v: \tilde{M} \longrightarrow \mathbb R$ (which is uniquely defined 
up to some additive constant) which is the solution of
\begin{equation}
\partial_z  (u - i  \, v ) =0  \qquad  \qquad (\mbox{and hence} \quad \partial_{\bar z}
(u +   i \, v ) =0 \, ) \, .
\label{eq:1.1}
\end{equation}
And we set 
\[
U: = u + i \, v \, .
\]
Recall that $U$ is a holomorphic function from $\tilde M$ into $\mathbb C$. 
The property that $u$ takes positive values in $M$ and vanishes on $\partial M$ 
can be translated into the fact that $U$ maps $\tilde{M}$ to 
\[
\mathbb C^+ := \{ w  \in \mathbb C \, : \,  \Re \, w  > 0\} \, ,
\]
and $\partial \tilde{M}$ to $i \, \mathbb{R}$. Since $\Phi 
\neq 0$ on $\tilde{M}$ there exists a unique holomorphic function $h$ on 
$\tilde{M}$ such that 
\[
dU =  \partial_z U \, dz = h \, \Phi \, .
\] 
We deduce from the fact that $u$ vanishes on $\partial \tilde M$ and from 
(\ref{eq:nblu1}) that $\nabla_n U  = 1$, if $n$ denotes the inward 
unit normal vector to $\partial \tilde M$, and hence
\begin{equation}
\| \partial_z U \|_g =  1 \qquad \mbox{on} \qquad \partial \tilde M \, .
\label{eq:zW}
\end{equation}
Now, condition (\ref{eq:nblu1}) translates into the fact that
\[
\|\Phi \|_g =  \|dF \|_g = 1 = \|dU \|_g \, ,
\]
on $\partial\tilde{M}$. Clearly, this is  equivalent to the fact that
\[
| h |= 1\qquad \hbox{on } \qquad \partial \tilde{M} \, .
\]
Therefore, we end up with the following data~:
\begin{enumerate}
\item[(i)] An oriented simply connected complex surface $\tilde{M}$ with 
smooth boundary $\partial \tilde{M}$. \\
\item[(ii)] A holomorphic function $U$, defined on $\tilde{M}$, which 
takes values in $\mathbb{C}^+$ and which maps $\partial \tilde{M}$ 
into $i \, \mathbb{R}$. \\
\item[(iii)] A holomorphic function $h$, defined on $\tilde{M}$, such that $|h|=1$ 
on $\partial \tilde{M}$ and for which the $1$-form $\Phi$ defined by  $\Phi : = 
\frac{1}{h} \, d U$ does not vanish on $\tilde{M}$. \\
\end{enumerate}
By analogy with the theory of minimal surfaces, we call these data the Weierstrass 
type representation formula for {\em exceptional flat surfaces}.
 
Conversely, given a set of such data, we can define the map 
$F : \tilde{M} \longrightarrow \mathbb{C}$ by integrating $dF = \Phi$. 
Thanks to (iii), this map is an immersion and its image is an immersed 
{\em exceptional flat surface} with {\em roof} function given by
\[
u  = \Re \, U \, .
\]
In the next section, we will give some explicit examples of such 
constructions when $\partial \tilde{M}$ is equal to $\partial D\setminus 
\{\alpha_1,\ldots, \alpha_n\}$, where $\alpha_1,\ldots, \alpha_n$ is a finite collection 
of points on $\partial D = S^1$.

\begin{example} We illustrate this Weierstrass type formula  
by giving some (rather pathologic) example.  We consider $M = \mathbb C^+$, 
the function $U(z) =z$ and 
\[
F(z) = \int_0^z e^{- \sinh \zeta} \, d\zeta \, .
\]
Note that $\partial_z F$ is $2 i \pi$-periodic and this implies that 
$F(z+2 i \pi) = F(z) +C$, where the constant $C$ is given explicitly by
\[
C : = i \, \int_0^{2\pi}e^{- i\sin s} \, ds \, .
\]
Moreover we observe that, for $x>0$, 
\[
F(x+i y) = F(i y) + \int_0^x e^{- \sinh (s+iy)} \, ds \, ,
\]
converges to $+\infty$ as $x \rightarrow +\infty$ if $y=0$, but this quantity is bounded 
if $|y - \pi|  < {\pi\over 2}$ and even admits a finite limit as $x \rightarrow +\infty$.

Hence, in addition to the {\it regular boundary} $F(i\, \mathbb{R})$ (which is a smooth periodic curve), 
the image of $F$ has a singular boundary which is the set of points which are the limits 
$\lim_{u\rightarrow +\infty}F(x+i \, y)$, as $u$ tends to $+\infty$, for the values of $y$ for 
which this limit exists. The {\em roof} function tends to infinity along this singular boundary.
\end{example}

\section{Examples of exceptional flat surfaces}
\setcounter{equation}{0}

Thanks to the Weierstrass type representation described in the
previous section, we can give many nontrivial examples of {\em
exceptional flat surfaces}. We keep the notations introduced in the 
previous section.

The construction makes use of an integer
$n \in \mathbb N\setminus \{0\}$ and the Riemann surface $D = \{ z
\in \mathbb C \, :  \, |z| < 1 \}$. On $D$, we define the
holomorphic functions
\[
h(z) = z^{n-1} \, ,
\]
and 
\[
U(z): = \frac{1+z^n}{1-z^n} \, .
\]
Then, the $1$-form $\Phi$ is given by 
\[
\Phi (z)  :=   \frac{2n}{(1-z^n)^2} \, dz \, ,
\]
Observe that both $U$ and $\Phi$ have singularities at the $n$-th roots of unity. 
The function $F$ is then obtained by integrating $\Phi$ and the {\em roof} function 
$u$ is then defined by $u =  \Re U$.

\begin{itemize}

\item[(i)] When  $n=1$, we can take
\[
F(z) = \frac{1+z}{1-z} \, .
\]
In this case, we simply have $F(D) = \mathbb C^+$ and we recover
the fact that the half plane is an {\it exceptional domain}. This
exceptional domain is the counterpart of the plane
in the framework of minimal surfaces. \\

\item[(ii)]  When $n= 2$,  we can take
\[
F(z) = \frac{2z}{1-z^2} + \log \left(  \frac{z+1}{z-1} \right) \, .
\]
In this case, the {\em exceptional flat surface} we find can be 
isometrically embedded in $\mathbb C$ and hence $F(D)$
is an {\it exceptional domain}. In fact,  $F(D)$ corresponds (up to some
similarity) to the domain $\Omega$ which has been defined in 
Proposition~\ref{pr:2.1}. This exceptional domain is the counterpart 
of the catenoid in the 
framework of minimal surfaces. \\

\item[(iii)] Finally, when $ n \geq 3$ the {\it exceptional flat surface} we find
cannot be isometrically embedded in $\mathbb C$ anymore. They are
the counterpart, in this setting, of the minimal $n$-noids described
in \cite{Jor-Mee}.\\

\end{itemize}

Let us analyze this example further. The function $U$ can be written
as
\[
U(z) = - \frac{1}{n} \,  \sum_{k=1}^{n} \frac {z + \alpha^{k} }{z - \alpha^{k}}  \, ,
\]
where $\alpha : = e^{i2\pi/n}$. In particular, $\Re \, U$ is nothing
but  a multiple of the sum of the Poisson kernel on the unit disc
with poles at $1, \alpha, \ldots, \alpha^{n-1}$. Next, observe that
\[
dU = z^{n-1} \, \frac{2n}{(1-z^n)^2} \, dz \, ,
\]
so that  the function $h$ is cooked up to counterbalance the zero 
of $dU$ and  ensure that $\Phi$ does not vanish in the unit disk, while
keeping the condition $|dU|^2 = |\Phi|^2$ on $\partial D$.

This example can be generalized as follows~:  Consider $n$ 
distinct points $\alpha_1, \ldots, \alpha_n \in S^1 \subset \mathbb C$ 
and $a_1, \ldots, a_n > 0$. We define
\[
U(z): = - \sum_{k=1}^n  a_k \, \frac{z+\alpha_k}{z-\alpha_k} \, .
\]
It is easy to check that $\Re \, U$ is positive (since each function 
\[
z\longmapsto - \displaystyle \frac{z+\alpha_k}{z-\alpha_k}\, ,
\]
maps $D$ to $\mathbb{C}^+$)  and vanishes on
$\partial D \setminus \{\alpha_1, \ldots, \alpha_n\}$. We have
\[
\prod_{k=1}^n (z-\alpha_k)^2 \, dU  = P(z) \, dz \, ,
\]
where $P$ is a polynomial which depends on the choice of the 
points $\alpha_1, \ldots, \alpha_n$ and the weights $a_1, \ldots, 
a_n$. Let us assume that $P$ does not vanish on $\partial D$ 
and let us denote by $z_1, \ldots, z_\ell$ the roots of $P$, counted 
with multiplicity) which belong to the unit disc. We
simply define
\[
h(z)  := \prod_{j=1}^\ell \, \frac{z \, - z_j}{z \, \bar z_j - 1}\,  ,
\]
and the $1$-form $\Phi$ by $ \Phi : =  \frac{1}{h} \, dU$. Integration 
of $\Phi$ yields a $2n$ dimensional  family of {\em exceptional flat 
surfaces} which are immersed in $\mathbb C$.

\section{A global  Weierstrass type representation}
\setcounter{equation}{0}

In this section, we show  that {\em exceptional flat surfaces} whose
immersion in $\mathbb{C}$ have finitely many {\em regular ends} and 
are locally finite coverings of $\mathbb{C}$  are precisely the examples 
presented in the previous section. We use the notations introduced 
in section \S \ref{preglobal} and we set 
\[
\widehat{M}: = M \cup\partial M \, .
\] 
We further assume that $M$ is \emph{simply connected} and that 
$\partial M\neq  \emptyset$. In particular, $M$ has the conformal type 
of  the unit disk $D$, and without loss of generality, we can assume that 
$M$  is indeed equal to $D$ and consider $\bar D$ as a natural 
compactification of $M$. We denote by $F$ an orientation 
preserving, holomorphic, isometric immersion $F: (\widehat{M}, g) \longrightarrow (\mathbb{C},
g_\mathbb{C})$. Recall that
\[
\|dF\|_g = 1\, ,
\]
on $\partial M$. Some natural  {\em hypotheses} will be 
needed~: 

\begin{itemize}
\item[(H-1)]  {\em $M$ has finitely many ends.} This means that
\[
\partial M = \partial D \setminus \cup_{j=1}^n E_j  = \cup_{j=1}^n I_j \, ,
\] 
where each $E_j \subset S^1$ is a closed arc and $I_j \subset S^1$ is an open arc. \\

\item[(H-2)]  {\em $F$ is proper.} This means that $F (w)$ tends to infinity as $w$ tends to $\cup_{j=1}^n E_j$.\\

\item[(H-3)]  {\em Each end of $N$ is regular.} This means that
the image of $I_j : =  (\theta_j^-, \theta_j^+)$ by $F$ is a curve 
$\Gamma_j$ which is asymptotically parallel to fixed
directions at infinity. In other words, there exist two unit vectors
$\tau_j^-$ and $\tau_j^+ \in S^1\subset \mathbb{C}$ such that
\[
\lim_{ \theta \in I_j \, , \, \theta \rightarrow \theta_j^\pm} {F(e^{i\theta}) 
\over |F(e^{i\theta})|} = \tau_j^\pm \, .
\]
Observe that this is for example the case if we assume that $\Gamma_j$ have finite total curvature. 

\item[(H-4)] {\em The mapping $F$ is a locally finite covering.} This means that 
there exists $d\in \mathbb{N}^*$ such that, for any $z\in \mathbb{C}$, the
cardinal of $\{\zeta \in M  \, : \,  F(\zeta) = z \}$ is less than or equal to
$d$.\\

\end{itemize}

The main result of this section reads~:
\begin{theorem}
Assume that $M$ be a simply connected {\em exceptional flat surface}  
and let $F : M \longrightarrow \mathbb{C}$ be an isometric immersion. 
Further assume that (H-1), \ldots , (H-4) hold and we identify $M$ with 
$D$. Then, there exist $\mu \in \mathbb R$, $n$ distinct points $\alpha_1,
\ldots, \alpha_n \in S^1$ and $n$ constants $a_1,\ldots , a_n >0$ such 
that
\[
dF = e^{i \, \mu} \, \prod_{k=1}^m{\overline{z_k} \, z-1\over z - z_k} \, dU.
\]
where $z_1,\ldots, z_m \in \bar D$ denote the zeros (counted with 
multiplicity) of $dU$ where $U$
\[
U( z ) : = - \sum_{j=1}^n a_j \, \frac{z + \alpha_j}{ z- \alpha_j}.
\]
in $\bar D$.
\label{theoremglobal}
\end{theorem}
The proof of the Theorem is decomposed into the following Lemmas and 
Propositions. We start by analyzing the ends $E_j$ and show that they 
reduce to isolated points $\alpha_1, \ldots, \alpha_n$. Next, we analyze 
the behavior of $F$ near the points $\alpha_j$ and show that $F$ does not
have any essential singularity there. Then, we proceed with the analysis  of
the function $U$ and show that it has the expected form. The proof of the Theorem 
is completed with the study of the  function $h$.

As promised, we first analyze the sets $E_j$. This is the contain of the 
following~:
\begin{lemma}\label{pointsalinfini} Under the assuptions of Theorem~\ref{theoremglobal},  
there exists a finite number of points $\alpha_1,\ldots , \alpha_n \in \partial D =S^1$ 
such that $\widehat{M} = \bar D \setminus\{ \alpha_1,\ldots ,\alpha_n\}$.
\end{lemma}
\begin{proof}  We need to show that each interval $E_j$ is reduced to a point.
This essentially follows from the fact that we can prove that the capacity of $E_j$ 
vanishes.

We argue by contradiction and suppose that, for some $j$, $E_j$ is
an arc of positive arc length. This implies that we can
find some $ \ell \in (0, \pi/2)$ and some arc $E \subset E_j$ of length 
$\ell$. Our problem being invariant under the action of homographic 
transformation of the unit disk, 
without loss of generality, we can assume that $E$ is the image of
$[-{\ell \over 2}, {\ell \over 2}]$ by $s \longmapsto e^{is}$ and, reducing $\ell$ if this is necessary, 
we can also assume that the opposite arc $-E$ (which is the 
image of $[- {\ell \over 2}, {\ell \over 2}]$ by $s \longmapsto - e^{is}$) is contained 
in $S^1 \setminus \cup_{j=1}^n E_j$. 

Recall that for any smooth function defined on $(a,b)$ which satisfies $f(b)=1$ and $f(a)=0$, we have 
\[
1 = f(b)-f(0) =\int_a^b f'(s) \, ds \leq \left( \int_a^b (f')^2 (s) \, ds\right)^{1/2} \, \sqrt{b-a} \, .
\]
If in addition, $b-a \leq 2$, we conclude that 
\[
\int_a^b (f')^2 (s) \, ds \geq \frac{1}{2}Ê\, .
\]
Now, assume that we are given a smooth function $f : \overline{D} 
\longrightarrow \mathbb{R}$
such that $f = 1$ on $E$ and $f = 0$ on $-E$, using the previous inequality, we can write
\begin{equation}
\label{endessous}
\int_{D} \|\nabla f \|_{g_{\mathbb C}}^2 \, dx\, dy \geq \int_{D\cap \{|x|< \sin \ell/2\}}
 |\partial_y f|^2 \, dx\, dy \geq \int_{|x| \leq \sin (\ell/2) } {1\over 2} \, dx = \sin (\ell/2) \, .
\end{equation}

Given $ R > r >0$ we let $\chi : \mathbb{C} \longrightarrow \mathbb{R}$ be defined by
\[
\chi (z) = \left\{ 
\begin{array}{cll}
0 & \qquad \hbox{if } \qquad |z| \leq r \\
\displaystyle \frac{\log \frac{|z|}{r}}{\log \frac{|z|}{R}} & \qquad \hbox{if } \qquad r \leq  |z| \leq  R \\
1 & \qquad \hbox{if } \qquad  R \leq |z| \, ,
\end{array}\right.
\]
and we define $f: D \longrightarrow \mathbb{R}$ by $f: = \chi \circ F$. Since $F$ is conformal, we can write
\[
\int_{D} \|\nabla f \|^2_{g_{\mathbb C}} \, dx \, dy   = \int_{D} \|\nabla f \|^2_g \, dvol_g  \, .
\]
Now, using (H-4), we conclude that 
\begin{equation}
\int_{D} \|\nabla f \|^2_g \, dvol_g  \leq d \int_{\mathbb C} \|\nabla \chi \|^2_{g_{\mathbb C}} \, dx\, dy  = d \, \frac{2\pi}{\log \frac{R}{r}} \, .
\label{frer}
\end{equation}

Fixing $r >0$ large enough, we can ensure that $f$ is identically equal to $0$ on $-E$. Using (H-2), we see that $f$ is identically equal to $1$ on each $E_j$, and in particular on $E$. Therefore, $f$ can be used in (\ref{endessous}) which together with (\ref{frer}) yields
\[
2 \, \pi \, d \geq \sin (\ell/2) \, \log \frac{R}{r} ,
\]
independently of $R >r$. Letting $R$ tend to infinity, we get a contradiction and the proof of the result is complete. \end{proof}

Therefore, we now know that 
\[
E_j : = \{\alpha_j\} \, .
\]
Without loss of generality we can assume that $\alpha_1, \ldots, \alpha_n$ are arranged counterclockwise along $S^1$. We agree that $\alpha_0 := \alpha_n$ and $\alpha_{n+1} : = \alpha_1$ and that, for each $j=1, \ldots , n$, the arc $I_j$ is positively oriented and joints $\alpha_j$ to $\alpha_{j+1}$. We now analyze the singularities of $F$ close to $\alpha_j$.

Given $j =1, \ldots, n$, we denote by $S (\alpha_j, r)$ the circle of radius $r >0$ centered at $\alpha_j$. We define  
\[
\gamma_j := \bar D \cap S (\alpha_j, r) \, .
\]
which we assume to be oriented clockwise. The {\em angle} $\theta_j \in \mathbb R$ at $\alpha_j$ is 
defined by
\[
\theta_j : = - \lim_{r\rightarrow 0} \int_{\gamma_k} F^*d\theta \, ,
\]
where $d \theta : =  \Im \frac{dz}{z}$.
Observe that, thanks to (H-3), $\theta_j$ is well defined  and we have 
\[
\tau_j^- = e^{i\theta_j} \, \tau_{j-1}^+ \, .
\]

With these definition in mid, we prove the
\begin{lemma}\label{comportementdeF} Under the assumption of Theorem~\ref{theoremglobal}, the function 
\[
 H_j (z) : =   (z- \alpha_j)^{\theta_j /\pi } \,  F(z) , 
\]
is holomorphic in a neighborhood of $\alpha_j$ in $\bar D \setminus \{\alpha_j\}$ and $H_j(\alpha_j) \neq 0$. 
\end{lemma}
\begin{proof} Without loss of generality, we can assume that $\alpha_j =1$. By right composing $F$ with the conformal transformation $z\longmapsto {1-z \over 1+z}$, we can replace $D$ by $\mathbb{C}^+$. Now, we define 
\[
G(z) := F(z)^{-\pi / \theta_j}
\] 
Observe that $G(0) =0$ by (H-2). Moreover, (H-3) together with the definition of $\theta_j$ implies that  the image by $G$  of a neighborhood of $0$ in $i \, \mathbb R$ is a $\mathcal C^1$-curve (and hence analytic). In particular, there exists some conformal transformation $T$ such that, for some $r > 0$, the image by $T \circ G$ of $i \, (-r, r)$ is a straight line segment in $i \, \mathbb{R}$. Then, it is possible to extend $T\circ G$ into some function $\tilde G$ which is defined on some neighborhood of $0$ in $\mathbb C$ by setting $\tilde G (z) = T (G(z))$ when $\Im \, z \geq 0$ and 
\[
\tilde G (z): =  - \overline{T (G(-\overline{z}))} \, ,
\]
when $\Im \, z \leq 0$.  The resulting function $\tilde G$ in bounded in a neighborhood of $0$ in $\mathbb C$ and  holomorphic away from $0$. It is well known that the singularity is then removable and  hence it is holomorphic and hence $\tilde G$ is actually holomorphic in a neighborhood of $0$. In particular, we can write
\[
G(z) = z^k \, H(z) \, ,
\]
close to $0$ where $H$ is a holomorphic function which  does not vanish at $0$. Going back to the definition of $G$, this implies that 
\[
F(z) =  (z - \alpha_j)^{- k \, \theta_j/\pi} \, H_j(z)
\]
where $H_j$ is holomorphic in a neighborhood of $\alpha_j$ and does not vanish at $\alpha_j$. But, the definition of $\theta_j$ readily implies that $k=1$.  This completes the proof of the result.
\end{proof}

As a corollary, we conclude that
\begin{equation}
H(z)  : = F(z) \, \prod_{j=1}^n (z- \alpha_j)^{\theta_j/\pi} \, , 
\label{corollairesurF}
\end{equation}
is a bounded holomorphic function in $\overline{D}$. Moreover, since $F$ tends to infinity as $z$ approaches $\alpha_j$, this implies that $\theta_j >0$.

We now make use of the fact that $M$ is an {\em exceptional domain} and hence there is a {\em roof} function $u: \widehat{M} \longrightarrow [0,+\infty)$ and we can define the holomorphic function $U:= u+i \, v$, where $v: \widehat{M} \longrightarrow\mathbb{R}$ is the (real valued) harmonic conjugate of $u$. The purpose of the next result is to show that $U$ is precisely given by the formula used in section\S 4.

\begin{lemma}\label{lemmesurU} 
Under the assumptions of Theorem~\ref{theoremglobal},  there exist $n$ constants $a_1, \ldots, a_n >0$ such that
\[
U(z) = -\sum_{j=1}^n  a_j \, \frac{z + \alpha_j}{z-\alpha_j} \, .
\]
\end{lemma}
\begin{proof} 
We first observe that it is possible to extend the function $U$ to all $\mathbb C\setminus  \{ \alpha_1,\ldots , \alpha_n\}$ by defining $V$ to be equal to $U$ in $\overline{D}\setminus \{ \alpha_1,\ldots , \alpha_n\}$ and 
\[
V(z)  := - \overline{U( 1/ \overline{z})} \, ,
\]
when $z \in \mathbb{C}\setminus\overline{D}$. The key observation is that, since $\Re \, U = 0$ on $\partial D\setminus \{ \alpha_1,\ldots , \alpha_n\}$, $V$ is continuous and in fact holomorphic on $\mathbb{C}\setminus \{ \alpha_1,\ldots , \alpha_n \}$. Moreover
$V$ converges to $V(\infty):= -\overline{U(0)}$ at infinity.

We proceed with the proof that the function $V$ has no essential singularity at any $\alpha_j$, 
this is a simple consequence of Picard's big Theorem. By definition, $\Re\, V$ vanishes on $I_j$ 
and is positive in $D$. Therefore, the outward normal derivative of $\Re\, V$ on $I_j$ is negative. 
This implies that the tangential derivative of $\Im\, V$ on 
$I_j$ does not vanish and hence that $\Im\,V$ is strictly monotone on each $I_j$. This 
shows that there exists some neighborhood $\mathcal V$ of $\alpha_j$ in 
$\mathbb{C}$ such that any element of $i \, \mathbb{R}$ is achieved by $V$ 
at most  twice in $\mathcal V$  (that is, at most once on $I_j$ 
and at most  once on $I_{j-1}$,  and certainly not in $\mathcal V
\setminus \partial D$, since $V$ takes values in $\mathbb{C} \setminus i 
\, \mathbb{R}$ away from $\partial D$). Picard's big Theorem \cite{conway} then 
implies that $\alpha_j$ is not an essential singularity of $V$. Hence $\alpha_j$ 
is either a removable singularity of $V$ or a pole. 

Since $\|\nabla u\|_g \equiv 1$ on $\partial M$, this forces $ | \partial_z U | 
 = |\partial_z F|$ on $\partial M$, and since $|\partial_z F|$ tends to $+\infty$ 
 at $\alpha_j$ so does $|\partial_z U|$ and hence all $\alpha_j$ are poles 
 of $V$.

We are now interested in the zeros of $V$. Since $\Re\,V$ takes positive values 
in $D$ and negative values in $(\mathbb{C}\cup \{\infty\})\setminus \overline{D}$, 
we know already that the only possible zeros of $V$ are on $\partial D$. 
Moreover, we have already seen that, along $I_j$,  the function $V = i \, v$ 
where  $v$ is strictly monotone. Furthermore since $\alpha_{j-1}$ and $\alpha_j$ 
are poles of $V$, $|V|$ must converges to $+\infty$ when one tends to 
$\alpha_{j-1}$ or $\alpha_j$. Because of the continuity of $v$ along each $I_j$ 
then implies that $v$ vanishes exactly at one point $\beta_j$ on each $I_j$. Moreover, 
this zero is simple, since if it would be a zero of order $k>1$, this would imply that 
the zero set  of $\Re\,V$ near $\beta_j$ contains $k$ curves intersecting at $\beta_j$. This 
would then force $\Re\,V = \Re \, U$ to vanish in $D$, which is in 
contradiction with our hypothesis.

Finally, we prove that $V$ has only simple poles. We know that $V$ extends 
meromorphically to a map on $\mathbb{C}P^1 = \mathbb{C} \, \cup \, \{\infty\}$ with 
no pole nor zero at infinity. Furthermore, $V$ has exactly $n$ simple zeros and 
$n$ poles, hence these poles must be simple. To summarize, the function $V$ 
can be written as a linear combination of the constant function and functions of 
the form $z \longmapsto \frac{1}{z- \alpha_j}$. Without loss of generality, this amounts 
to say that $V$ can be written as 
\[
V(z) =  a - \sum_{j=1}^n   a_j \, \frac{z+ \alpha_j}{z-\alpha_j} \, ,
\]
where $a$ and the $a_j$ are complex numbers. Using the fact that, 
by construction, $V( 1 / \overline{z} ) = - \overline{V(z)}$, we conclude that 
$ a \in i \, \mathbb R$ and also that $a_j \in \mathbb R$. Moreover, $\Re \, U$ 
being positive, this implies that the $a_j$ are positive real numbers. This 
completes  the proof of the result since $U$ is defined up to the addition of 
some element of $iÊ\, \mathbb R$. 
\end{proof}

We  are now in a position to complete our analysis of the function $F$. Since $F$ 
is an immersion $dF\neq 0$ on $\widehat{M}$. Hence there exists a unique 
holomorphic function 
$h$ on $\widehat{M}$ such that
\begin{equation}
\label{du=hdF}
\partial_z U = h \,  \partial_z F \, ,
\end{equation}
on $\widehat M $. Moreover, since $\|\nabla u \|_g \equiv 1$ on $\partial M$, this implies 
that $ | h | \equiv 1$ on $\partial M$.  In the next result, we analyze the function $h$, this 
will complete the proof of Theorem~\ref{theoremglobal}.

\begin{lemma}
\label{lemmesurH}
Under the assumptions of Theorem~\ref{theoremglobal}, there exists a constant 
$e^{i \, \mu} \in \mathbb R$ such that the function $h$ defined by (\ref{du=hdF}) 
has the form
\[
h(z) = e^{-i\, \mu}\prod_{k=1}^m\frac{z-z_k}{\overline{z_k}z-1}.
\]
where $z_1,\ldots z_m$ are the zeros of $\partial_z U$ in $D$ counted with multiplicity.
\end{lemma}
\begin{proof} The function $h$ is holomorphic in $D$ and satisfies $|h| = 1$ on $\partial D 
\setminus \{\alpha_1,\ldots, \alpha_n\}$. We can extend $h$ as a holomorphic function $H$ 
which defined on $(\mathbb{C} \cup \{ \infty \}) \setminus \{ \alpha_1, \ldots , \alpha_n \}$ 
by setting $H(z) := h(z)$ for all $z \in  \overline{D}\setminus \{ \alpha_1,\ldots ,\alpha_n\}$ 
and 
\begin{equation}
\overline{H(z)} := \frac{1}{{h(1/\overline{z})}} \, ,
\label{eqh}
\end{equation}
for all $z \in \mathbb{C}\setminus\overline{D}$. Clearly $H$ is locally bounded in 
$\overline{D} \setminus \{ \alpha_1,\ldots , \alpha_n\}$, its only singularities in $(\mathbb{C}\cup\{\infty\})
\setminus \overline{D}$ are poles which are the images by $z\longmapsto 1/\overline{z}$ 
of the zeros of $h$ and hence is meromorphic outside $\{ \alpha_1,\ldots , \alpha_n\}$. 
But, Lemma \ref{comportementdeF} and (\ref{corollairesurF})  imply that, near $\alpha_j$, 
$|H|$ is bounded by a constant times $ | z - \alpha_j |^{-k_j}$ for some $k_j >0$. Therefore, 
$\alpha_j$ is not an essential singularity of $H$ and hence, $H$ is meromorphic in 
$\mathbb{C} \cup \{ \infty \}$.  

Observe that $|H(z)|=1$ on $\partial D \setminus \{ \alpha_1, \ldots , 
\alpha_n\}$  and this implies that the points $\alpha_j$ are not poles of $H$. Therefore, the 
singularities $\alpha_j$ of $H$ are removable.  Also, we have 
\[
\Delta \, |H|^2 =  4 \partial_z \partial_{\bar z} \, |H|^2 =  4 \, |\partial_z H|^2 \geq 0 \, ,
\]
and since $|H| =1$ on $\partial D$,  the maximum principle implies that  $|H|\leq 1$ in $D$.

Now, $H$ being bounded in $\overline{D}$, it does not have poles in this set and this also
implies that  $H$ has no zeroes in $(\mathbb{C}\cup\{\infty\}) \setminus D$ (because otherwise 
$H$ would have poles in $\overline{D}$ by (\ref{eqh}). Therefore, if $z_1,\ldots, z_m \in D$ denote 
the zeros of $H$ (counted with multiplicity), then the poles of $H$ are given by $1/\overline{z_1} ,
\ldots ,1/\overline{z_m}$ (also counted with multiplicity). It is then a simple exercise to check that $H$ is of the form 
\[
H(z) = C \, \prod_{k=1}^m \frac{z - z_k}{ \overline{z_k}  \, z-1} \, ,
\]  
for some constant $C \in \mathbb{C}$. Finally, the condition that $|H(z)| = 1$ on $\partial D$
 forces $|C| = 1$. This completes the proof of the result.
\end{proof}

\section{A Bernstein type result for $2$-dimensional exceptional domains}
\setcounter{equation}{0}

We prove the following Bernstein type result for $2$-dimensional {exceptional domains}~:
\begin{proposition}
Assume that $\Omega$ is a $2$-dimensional {\em exceptional domain} which is conformal
 to $\mathbb C^+$ and let $u$ be a {\em roof} function on $\Omega$. We further assume that 
 $\partial_x u > 0$ in $\Omega$, then $\Omega$ is a half plane.
\end{proposition}
\begin{proof}  Since we have assumed that $\Omega$ is conformal to $\mathbb C^+$, 
there exists a holomorphic map $\Psi : \mathbb C^+ \longmapsto \Omega$. We then define 
\[
H  := (\partial_z u) \circ \Psi \, .
\]
The function $H$ is holomorphic in $\mathbb C^+$ and does not vanish (since we have 
assumed that $\partial_x u \neq 0$). Moreover, $|H| \equiv 1$ on $\partial  \mathbb C^+$. 
We can write
\[
H = e^{i \Theta} \, ,
\]
where $\Theta$ is a holomorphic function defined in $\mathbb C^+$ which is real valued 
on the imaginary axis.  This means that 
\[
\Im \, \Theta =0 \quad \mbox{when} \quad \Re \, z = 0 \, .
\]
Since we have assumed that $\partial_x u >0$, we also conclude that $\Re \, \Theta \in 
(-\pi/2, \pi/2)$. 

We can extend $\Theta$ as a holomorphic function $\Tilde \Theta$ in $\mathbb C$ as follows~: for 
all $z \in \mathbb C$ such that $\Re \, z \geq 0$ we set
\[
\tilde \Theta (z)  := \Theta (z) \, ,
\]
while, when $\Re \, z <0$, we set
\[
\tilde \Theta (z)  : = \overline{\Theta (-\bar z)} \, .
\]
It is easy to check that $\tilde \Theta$ is a holomorphic function : in fact, the real 
part of $\Theta$ is extended as an even function of $\Re z$ while the imaginary 
part of $\Theta$ is extended as an odd function of $\Re z$. The fact that $\tilde 
\Theta$ is $\mathcal C^1$ is then a consequence of the fact that $\Im \, \Theta = 
0$ on the imaginary axis and the fact that $\Theta$ being holomorphic, $\partial_x 
\Re \, \Theta =0$ on the imaginary axis of $\mathbb C$.

Observe that the real part of $\tilde \Theta$ is a bounded harmonic function, and, 
as such, it has to be constant. The function $\tilde \Theta$ being holomorphic, we 
conclude that it is constant. But this implies that the gradient of $u$ is constant and 
hence the level sets of $u$ are straight lines. This implies that $u$ only depends 
on one variable and hence it is a affine function. This completes the proof of the 
result.
\end{proof}

As a Corollary, we also prove the~:
\begin{corollary}
There is no {\it exceptional domain} contained in a wedge
\[
\Omega \subset \{ z \in \mathbb C \, : \, \Re \, z \geq \kappa \, | \Im \, z | \} \, ,
\]
for some $\kappa >0$. 
\end{corollary}
\begin{proof} 
The proof is by contradiction. If $\Omega$ were such an {\it exceptional domain}, 
there would exist on  $\Omega$ a {\em roof} function $u$. One can apply the moving plane 
method \cite{Ser}, \cite{Gid-Ni-Nir} to prove that $\partial_x u > 0$ and hence 
that $\partial \Omega$ is a graph over the $y$-axis. Observe that, since $\Omega$ 
is contained in a half plane, there is no bounded, positive, harmonic function on 
$\Omega$ having $0$ boundary data on $\partial \Omega$ (otherwise one could 
use an affine function as a barrier to obtain a contradiction). Certainly, $\Omega \cup 
\partial \Omega$ is conformal to $\bar D\setminus E$ where $D$ is the unit disc and 
$E$ is a closed arc included in $S^1$. Necessarily, $E$ is reduced to a point since 
otherwise we can construct bounded, positive, harmonic functions on $E$ which 
have $0$ boundary data on $S^1 \setminus E$ and these would lift to bounded, 
positive, harmonic function on  $\Omega$, with $0$ boundary data, a contradiction. 
Therefore, we conclude that $\Omega$ is conformal to $\mathbb C^+$. The assumptions
of the previous Lemma are fulfilled and hence we conclude that $\Omega$ is a half plane, 
which  is a contradiction. 
\end{proof}

\section{Open problems}
\setcounter{equation}{0}

We have no non trivial example of {\it exceptional domain} in higher dimensions,  
$\mathbb R^m$, for $m \geq 3$ (beside the examples described in section 2). 
In dimension $m=2$, it is tempting to conjecture that (up to similarity) 
the only {\it exceptional domains} which can be embedded in $\mathbb 
R^2$ are the half spaces, the complement of a ball and the example 
discussed in section 2.

\end{document}